\documentclass[sn-mathphys,Numbered,iicol]{style}

\usepackage{graphicx}%
\usepackage{multirow}%
\usepackage{amsmath,amssymb,amsfonts}%
\usepackage{amsthm}%
\usepackage{mathrsfs}%
\usepackage[title]{appendix}%
\usepackage{xcolor}%
\usepackage{textcomp}%
\usepackage{manyfoot}%
\usepackage{booktabs}%
\usepackage{algorithm}%
\usepackage{algorithmicx}%
\usepackage{algpseudocode}%
\usepackage{listings}%

\usepackage{enumitem}
\newlist{compactitem}{itemize}{3} 
\setlist[compactitem]{label=\textbullet,leftmargin=2em,nosep}

\usepackage{xurl}
\usepackage{booktabs}
\usepackage{multirow}
\usepackage{makecell}
\newcommand{\datatext}[1]{\texttt{#1}}
\newcommand{\suppurl}{https://takanori-fujiwara.github.io/s/adv-attack-on-vis}
\renewcommand{\quad}{\hskip0.5em\relax}

\setlist[description]{style=unboxed,labelindent=0em}
\usepackage{dblfloatfix}
\usepackage{fancyhdr}
\begin{document}

\title[Adversarial Attacks on Machine Learning-Aided Visualizations]{Adversarial Attacks on Machine Learning-Aided Visualizations}


\author*[1]{\fnm{Takanori} \sur{Fujiwara}}\email{takanori.fujiwara@liu.se}
\author[1]{\fnm{Kostiantyn} \sur{Kucher}}
\author[2]{\fnm{Junpeng} \sur{Wang}}
\author[3]{\fnm{Rafael M.} \sur{Martins}}
\author[1,3]{\fnm{Andreas} \sur{Kerren}}
\author[1]{\fnm{Anders} \sur{Ynnerman}}

\affil[1]{\orgname{Link\"{o}ping University}, \orgaddress{\country{Sweden}}}
\affil[2]{\orgname{Visa Research}, \orgaddress{\country{United States}}}
\affil[3]{\orgname{Linnaeus University}, \orgaddress{\country{Sweden}}}


\abstract{Research in ML4VIS investigates how to use machine learning (ML) techniques to generate visualizations, and the field is rapidly growing with high societal impact. However, as with any computational pipeline that employs ML processes, ML4VIS approaches are susceptible to a range of ML-specific adversarial attacks. These attacks can manipulate visualization generations, causing analysts to be tricked and their judgments to be impaired. Due to a lack of synthesis from both visualization and ML perspectives, this security aspect is largely overlooked by the current ML4VIS literature. To bridge this gap, we investigate the potential vulnerabilities of ML-aided visualizations from adversarial attacks using a holistic lens of both visualization and ML perspectives. 
We first identify the attack surface (i.e., attack entry points) that is unique in ML-aided visualizations. 
We then exemplify five different adversarial attacks.
These examples highlight the range of possible attacks when considering the attack surface and multiple different adversary capabilities. Our results show that adversaries can induce various attacks, such as creating arbitrary and deceptive visualizations, by systematically identifying input attributes that are influential in ML inferences. Based on our observations of the attack surface characteristics and the attack examples, we underline the importance of comprehensive studies of security issues and defense mechanisms as a call of urgency for the ML4VIS community.}




\maketitle

\thispagestyle{fancy}

\section{Introduction}

Visualizations play a critical role in depicting relationships or patterns within an underlying dataset such that analysts can effectively explore, interact, and communicate the data.
Recently, researchers are actively investigating various applications of using machine learning (ML) techniques when generating visualizations optimized for certain tasks (e.g., chart recommendation~\cite{zhou2021table2charts}, text-chart translation~\cite{cui2020texttoviz}, and interaction prediction~\cite{ottley2019follow}).
This approach is often referred to as ML for visualization, or ML4VIS for short~\cite{wang2022ml4vis}.
Despite the wide usage, we argue that the current research focus on ML4VIS is imbalanced. The majority of ML4VIS research mainly focuses on the \textit{benefits} these techniques provide~\cite{wang2022ml4vis,wang2022dl4scivis,wu2022ai4vis}, and consequently largely overlooks the \emph{security issues} these techniques introduce.

Critically understanding security issues related to ML4VIS requires a more holistic perspective that is both visualization- and ML-oriented. 
This dual perspective is necessary as the joint use of ML and visualization techniques may cause additional security problems that are not prevalent in each respective field. 
However, relevant research investigations on these matters are currently disconnected. 
For example, though some visualization research has explored the vulnerabilities of visualization approaches, the focus of these investigations targets more general visualization practices (e.g., the influence of the choice of the aspect ratio on the recognition of correlation patterns~\cite{correll2017black,mcnutt2020surfacing}). These insights do not consider the specific security issues that ML approaches introduce (e.g., manipulating data transformation when generating ML-aided visualizations). 
Similarly, though the ML community is intensively studying the vulnerability of ML models from adversarial attacks~\cite{goodfellow2018making,papernot2018sok}, these security vulnerabilities have not been studied with respect to visualization. 
Adversaries can then take advantage of loopholes that are unattended.
For instance, while direct human intervention is vital to prevent critical harm from ML mispredictions~\cite{heer2019agency,hamon2020robustness,shneiderman2020human}, this scenario is generally not applicable to ML4VIS. 
In ML4VIS, an adversary's goal could be to harm the end users by corrupting output visualizations that humans use for their judgments.
This lack of an interdisciplinary perspective can introduce security concerns that both communities are overlooking. 

This paper aims to convey the importance of systematic studies related to the security vulnerabilities of ML4VIS approaches. 
To reveal structural vulnerabilities of ML-aided visualizations, we first outline the attack surface (i.e., a set of entry points for adversaries) by referring to the ML4VIS pipeline introduced by Wang et al.~\cite{wang2022ml4vis}.
We then characterize the unique aspects of the attack surface of ML-aided visualizations.
For this characterization, we focus on ML-aided visualizations that incorporate neural networks (NNs), as NNs are rapidly gaining increased attention and wide usage~\cite{wang2022ml4vis,wang2022dl4scivis,wu2022ai4vis}.
To demonstrate the unique characteristics of the attack surface with real possible threats, we design concrete examples of adversarial attacks on representative state-of-the-art ML4VIS methods.
The results from these exemplary attacks highlight critical vulnerabilities in the current ML4VIS approach that can directly impact future applications.

Based on the insights from our literature review, characterization of the attack surface, and attack examples, we discuss the identified research gaps on the lack of studies on vulnerabilities specific to ML4VIS and propose a research agenda outlining future research directions from a security perspective.
Our research reflects the urgency of this matter as several government and official institutions (e.g., EU~\cite{euaiact}, UK~\cite{ukairegulation}, and US~\cite{usaibill}) have already introduced and developed ML security regulations. 
We similarly argue that it is important that visualization researchers also take action to minimize the risk of any harm that can stem from using ML in conjunction with visualization. 
In summary, our primary contributions are to be:

\begin{itemize}
  \item{characterization of the attack surface of ML-aided visualizations, which produces an indication of potential threats (\autoref{sec:threat-model});}
  \item{five examples of adversarial attacks using two representative ML-aided visualizations that (1) describe the attack strategies and (2) analyze the attacked results (\autoref{sec:attack-examples}); and}
  \item{a research agenda outlining the future research direction for ML4VIS from a security perspective (\autoref{sec:discussion}).}
\end{itemize}

\section{Background and Related Work}
\label{sec:related-work}

Addressing security issues in ML-aided visualizations requires a dual set of ML and visualization considerations. 
We describe relevant works related to adversarial attacks and vulnerabilities from ML and visualization research. 
We also discuss a general overview of ML-aided visualizations.

\subsection{Adversarial Attacks on Machine Learning Models and Defenses}
\label{sec:rw-attack-ml}

With ML techniques actively used in real-world settings, Dalvi et al.~\cite{dalvi2004adversarial} posits a critical research agenda that can address the issue of how ``the presence of an adversary actively manipulating the data [can] defeat the data miner''.
As deep NNs are rapidly being used in various domains, such as vision and speech recognition, a significant portion of ML research is devoted to addressing security issues in NNs. 
One notable early result is an \emph{adversarial example} designed for convolutional NN (CNN) models by Szegedy et al.~\cite{szegedy2014intriguing}.
They demonstrated that an adversarial example can be easily constructed by adding a human-imperceptible perturbation into an input image. 
This perturbation can readily cause image misclassification even with state-of-the-art CNN models. 
These adversarial examples pose critical issues in real-world settings where, for example, an adversary may craft a stop traffic sign that an autonomous car will misclassify as a yield sign~\cite{papernot2017practical}. Researchers have since studied a multitude of efficient and intense adversarial examples on CNNs.

Adversarial examples on CNNs can be generally categorized as either \textit{white-box attacks} or \textit{black-box attacks} depending on the adversary's knowledge about the model of interest.
When an adversary knows detailed information about a target CNN model (i.e., white-box attacks), they can efficiently construct adversarial examples by referring to the gradients of the model's loss function~\cite{goodfellow2015explaining,su2019one}.
In contrast, black-box attacks are when the adversary has limited information available on a target model. 
In these cases, the adversary generally has two strategies. First, an adversary can build a \emph{substitute model} that performs the same classification task as the target model. 
The substitute model can then be used to generate adversarial examples like in white-box attacks~\cite{szegedy2014intriguing}. 
Another common black-box attack strategy is to \emph{reverse engineer} a target model from the collected input-output relationships by sending queries to the model~\cite{papernot2017practical}. 
Research continues to highlight new attack methods that have grown and diversified for other NNs, such as recurrent NNs (RNNs) and graph NNs (GNNs)~\cite{xu2020adversarial}.

Besides crafting adversarial examples, \emph{data poisoning}~\cite{biggio2012poisoning} (i.e., adding malicious training data) is another effective way to corrupt the model as exemplified by the problematic tweets made by Microsoft's chatbot Tay~\cite{neff2016automation}.
Data poisoning can also be used to perform backdoor attacks. 
Backdoor attacks can make NN models behave maliciously only when models receive inputs stamped with a trigger crafted by the adversaries~\cite{chen2017targeted}. 
Multiple surveys~\cite{li2022backdoor,gao2020backdoor} highlight how backdoor attacks can be easily prepared with various approaches, such as sharing maliciously pre-trained models with others. 
The growing range of these attacks highlights the need for defense strategies.

Defense strategies against adversarial attacks vary but all present shortcomings. 
One straightforward way to protect against white-box attacks is gradient masking, which conceals the gradient information from adversaries~\cite{papernot2016distillation}. However, even after gradient masking, an adversary can perform a black-box attack with a substitute model.
Another defense strategy is to generate more robust NN models through adversarial training (i.e., training using artificially created adversarial examples)~\cite{goodfellow2015explaining,szegedy2014intriguing}.
NN models produced with adversarial training are still vulnerable to out-of-sample adversarial examples.
A third popular strategy focuses on input validation and preprocessing, such as applying statistical methods to detect abnormal inputs (e.g., PCA-based detection~\cite{rubinstein2009antidote}) and data compression to exclude the perturbations (e.g., JPEG compression~\cite{dziugaite2016study}).
However, this strategy is highly domain-specific and is not generalizable across domains~\cite{goodfellow2018making}.
We discussed only three possible defense strategies. 
We refer readers to the taxonomy by Papernot et al.~\cite{papernot2018sok} and multiple surveys and reports for more details on attack and defense strategies ~\cite{akhtar2018threat,goodfellow2018making,hamon2020robustness}).

Visual analytics approaches have also devoted to investigating the mechanism of adversarial attacks on NNs to establish better defense strategies~\cite{cao2021analyzing,das2020bulff,li2023visual}. 
However, there is a research void in studying the impact of adversarial attacks on the generation of visualizations.

\subsection{Vulnerabilities of Visualizations}
\label{sec:rw-attack-vis}

Visualizations can become obscure, misleading, and even deceptive as a consequence of poorly prepared data~\cite{song2019wheres}, problematic visual designs~\cite{lauer2020deceptive,pandey2015deceptive}, viewers' cognitive bias~\cite{ellis2018cognitive,xiong2022seeing}, or any combinations of these.
For example, a visualization with missing value imputations that is not suited for target analysis tasks can decrease viewers' performance~\cite{song2019wheres}.
Subtle skews in visualizations, such as 3D effects on pie charts, can lead to wrong conclusions~\cite{lauer2020deceptive}.
A viewer's belief in the existence of correlations between two variables (e.g., the numbers of guns and crimes) can also influence the cognition of the correlation strength~\cite{xiong2022seeing}.
The choice of a visual representation in itself can additionally lead to bias in estimating a user's confidence in the presented visual representations~\cite{helske2021can}.
McNutt et al.~\cite{mcnutt2020surfacing} provided a conceptual model that frames these visualization vulnerabilities along the visual analytics pipeline.

By exploiting these visualization vulnerabilities, adversaries can easily create malicious visualizations.
Correll and Heer~\cite{correll2017black} discussed a man-in-the-middle attack on visualizations: an attack from a malicious visualization designer who aims to distract communication between data and viewers.
Adversaries (i.e., designers) can intentionally break visualization conventions~\cite{kennedy2016work} to create visualizations with problematic designs. 
Our work argues that even when adversaries do not have such a strong capability of manipulating visualization designs, various practical attacks can be imposed on ML-aided visualizations (see \autoref{sec:threat-model} and \ref{sec:attack-examples}).

To the best of our knowledge, no existing work explicitly studied defense strategies against adversarial attacks on visualizations.
Existing  works targeted on detecting flaws in visualizations~\cite{chen2022vizlinter}, mitigating cognitive biases~\cite{fan2022annotating,wall2017warning}, and testing the robustness of visualizations~\cite{mcnutt2020surfacing}.
For example, Chen et al.~\cite{chen2022vizlinter} developed a linting tool to detect erroneous visual specifications. 
McNutt et al.~\cite{mcnutt2020surfacing} introduced metamorphic testing for visualization, which measures how strongly a change to the input data can perturb a visualization outcome. 
This current state of literature presents security issues that are being overlooked within the visualization community.

\subsection{Machine Learning-Aided Visualizations}
Existing works~\cite{wang2022ml4vis,wang2022dl4scivis,wu2022ai4vis} provide comprehensive surveys on ML4VIS related to visual analytics, information visualization, and scientific visualization.
Based on these surveys, the following facts emphasize the importance of needing future studies regarding adversarial attacks on ML-aided visualizations.
\begin{description}[labelsep=1em]
    \item[Increase of research interest in ML4VIS.]
    Research on ML4VIS is rapidly growing. 
    Until 2015, only a few ML4VIS-related papers were published annually. 
    Since then, the number of ML4VIS-related publications radically increased~\cite{wang2022ml4vis}.
    This trend indicates an increasing interest in, and need for, using ML for visualization. 
    
    \item[Existence of broad and critical attack targets.] 
    ML4VIS approach is employed for various tasks, including data processing for visualization, feature extraction, visualization generation, user behavior prediction, and interpretation support of visualized results. 
    Also, ML4VIS approaches are utilized in life-critical applications, such as 3D medical image analysis and reconstruction~\cite{ghahremani2022neuroconstruct,wang2020deeporgannet,stacke2021measuring} as well as cancer genome analysis~\cite{gramazio2018analysis}.
    Because visualization is a fundamental tool to communicate and analyze data, by attacking ML-aided visualizations, adversaries could significantly and negatively impact critical applications in areas such as business, politics, and medicine.
    
    \item[Potential exposure to immediate threats.]
    Many existing ML-aided visualizations are based on NN-based models, using multilayer perceptrons (MLPs), CNNs, RNNs, GNNs, autoencoders, and/or generative adversarial networks.
    Adversarial attack methods have been developed for all of these NNs~\cite{akhtar2018threat,xu2020adversarial}. 
    Therefore, ML-aided visualizations might be inherently exposed to potential threats.
\end{description}

\section{Attack Surface of Machine Learning-Aided Visualizations}
\label{sec:threat-model}

We analyze the attack surface (i.e., set of potential entry points for attacks) of ML-aided visualizations.
For this analysis, we use Wang et al.'s ML4VIS pipeline~\cite{wang2022ml4vis}, which is derived from an extensive survey on ML4VIS approaches, to review the potential processes, inputs, and outputs involved in ML-aided visualizations.

\subsection{ML4VIS Pipeline}
\label{sec:pipeline}

\begin{figure}[t]
    \centering
    \includegraphics[width=\linewidth]{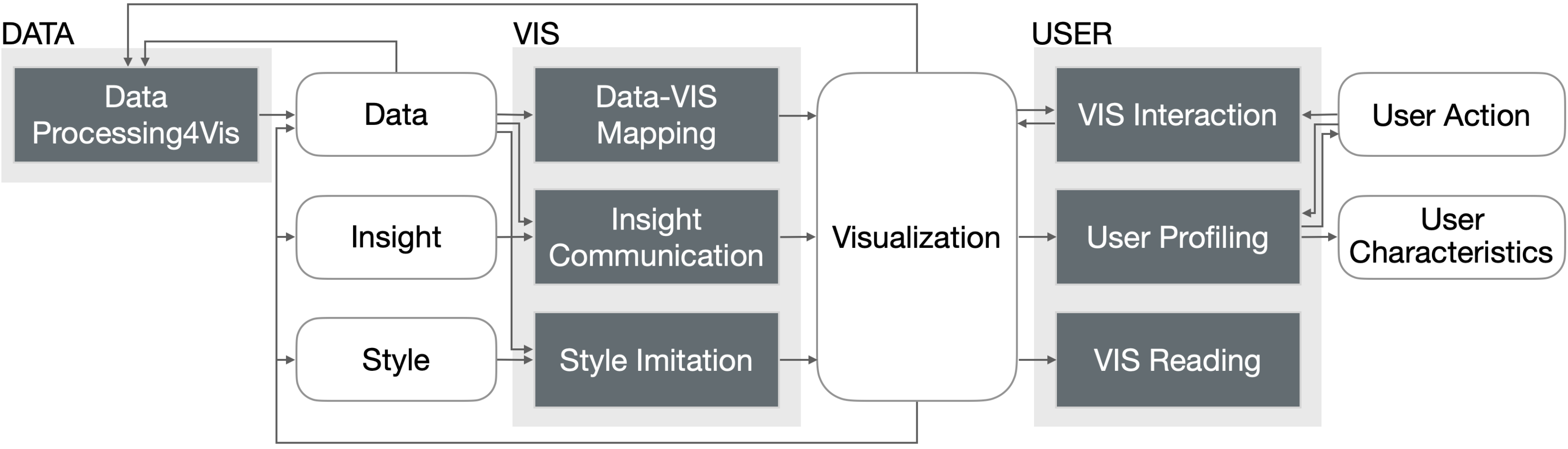}
    \caption{An ML4VIS pipeline introduced by Wang et al.~\cite{wang2022ml4vis}. This pipeline links seven visualization processes (gray blocks) that can benefit from ML. The processes are further categorized based on their main aim: processing \textit{data}, producing \textit{visualization}, or helping and understanding \textit{users}.}
    \label{fig:pipeline}
\end{figure}

As shown in \autoref{fig:pipeline}, Wang et al.'s ML4VIS pipeline consists of seven processes that can be individually aided by ML: Data Processing4VIS, Data-VIS Mapping, Insight Communication, Style Imitation, VIS Interaction, User Profiling, and VIS Reading.
Note that we consider that visualizations are ML-aided if they utilize ML for \textit{any} of these processes. 
ML does not have to be involved in all the processes.
For example, a user may utilize ML for data processing but still manually design the visual encodings to produce a visualization. 
Below, we briefly describe each process.

\textbf{Data Processing4VIS} prepares data for the subsequent visualization processes.
A typical ML method used for this process is dimensionality reduction (DR). 
This process can use raw data, existing visualizations, or both as input. 
For example, DR can utilize the previously generated visualization result to perform visually consistent updates~\cite{fujiwara2020incremental,rauber2016visualizing}.
It is worth noting that NN-based DR methods are becoming more actively developed for Data Processing4VIS~\cite{lai2022parametric,sainburg2021parametric,vandermaaten2009learning,hinterreiter2022paradime,zang2024dmtev}.

\textbf{Data-VIS Mapping} encodes processed data into visualizations by assigning visual marks and channels.
For this process, ML can recommend marks and channels suitable for a given dataset or accelerate the encoding process. 
These enhancements are often achieved by training NNs using supervised learning on labeled training data (e.g., data tables and their suitable visual marks)~\cite{wu2022multivision,zhao2022chartseer,zhou2021table2charts}.

\textbf{Insight Communication} aims to produce visualizations that effectively convey insights found in the data.
Wang et al.~\cite{wang2022ml4vis} distinguish this process from Data-VIS mapping based on whether the insights are used as inputs.
Because insights are frequently represented with texts, NN-based natural language processing can be applied to this process~\cite{cui2020texttoviz}.

\textbf{Style Imitation} reflects a given visualization's style in other visualizations. 
Visual styles include applied color palettes, chart decorations, etc.
Similar to Data-VIS Mapping, this process can be aided by NNs trained with supervised learning~\cite{tang2021plotthread,kwon2020deep}.

\textbf{VIS Interaction} takes user actions as input and then accordingly updates visualizations by using any of the aforementioned processes.
For this process, ML can aid in interpreting users' intentions as well as refine the interaction results.  
For example, NN-based classification models are used to improve the accuracy and speed of selecting visual elements (e.g., points in 2D scatterplots)~\cite{fan2018fast}.

\textbf{User Profiling} is to understand users from their action histories.
ML is applied to predict user behaviors (e.g., next interactions) or user characteristics (e.g., cognitive abilities, analysis goals, and personalities)~\cite{gramazio2018analysis,ottley2019follow}.
With such predictions, visualizations can provide better interaction latency, suggestions for the next actions, and desirable marks and channels for users and their analysis.

\textbf{VIS Reading} is the process of reviewing visualizations and understanding the encoded information. 
Through this process, the goal is to understand the visualized data, extract applied styles, and obtain insights. 
ML can help users automate VIS Reading, instead of relying on manual human inspections. 
For example, NNs can reconstruct input data corresponding to visualized results~\cite{poco2017reverse,ma2018scatternet}. 

As shown in \autoref{fig:pipeline}, these processes are intertwined via the inputs and outputs that are produced by or used for the other processes.
In addition, as with other visualization pipelines~\cite{wang2016survey}, there are likely iterative updates for each process based on newly added data and user actions.
We also want to note that the dataflow of this pipeline only reflects inputs and outputs at the \emph{inference} phase. 
For the \emph{training} phase, pre-generated outputs can be used as training data for supervised learning (e.g., Data-VIS mapping can be trained with a set of pairs of data and its visualization) for each process.

\subsection{Characterization of Attack Surfaces}
\label{sec:attack-surface}

We consider that all employed processes, inputs, and outputs in Wang et al.'s ML4VIS pipeline compose an attack surface given how adversaries may be able to manipulate inputs, corrupt ML/non-ML processes, and tamper with the output results.
Probable attack surfaces across different ML-aided visualizations reveal critical, unique characteristics in ML-aided visualizations:
\begin{description}[labelsep=1em]
    \item[C1. ML inputs and outputs specific to visualization.]
    Though existing ML-aided visualizations usually customize and utilize NNs developed from ML research~\cite{wang2022ml4vis,wang2022dl4scivis,wu2022ai4vis}, the inputs used for the training and inference phases are often specific to the visualization (e.g., \cite{kwon2020deep,ottley2019follow,ma2018scatternet}).
    Target outputs and loss functions to produce such visualization outputs can also be unique for ML-aided visualizations.
    For example, ML-aided visualizations often adapt NNs to analyze visualizations based on their visual marks and channels rather than pixel-based images~\cite{zhao2022chartseer,wu2022multivision}.
    Consequently, adversaries might design attack methods for ML-aided visualizations that are significantly different from those studied in the ML field, and existing vulnerability assessment and defense methods for NNs~\cite{papernot2018sok,goodfellow2018making} might not be as effective for such attacks.
    
    \item[C2. Exposure of ML outputs (and inputs) to adversaries.]
    Visualizations are usually intended to be reviewed by users. The processes within the ML4VIS pipeline, such as Data-VIS Mapping, Insight Communication, and Style Imitation, all produce visualizations as their outputs.
    Thus, when these processes employ ML, the visualizations are inference results that are abundant with information, which will inherently be observed by users as well as adversaries. 
    When the data generated by Data Processing4VIS is visualized without additional processes (e.g., directly visualizing DR results as 2D scatterplots~\cite{fujiwara2020incremental,hinterreiter2022paradime}), Data Processing4VIS also suffers from the same issue.
    Moreover, interactive visualizations often support details-on-demand~\cite{huang2008context}, enabling adversaries to access detailed information of raw or processed input data.
    This attack surface characteristic can enable adversaries to gain further knowledge of ML models (i.e., contributing to the \textit{reconnaissance} stage of cyberattacks~\cite{almohannadi2016cyber}).
    In contrast, C1 provides new opportunities to create attacks specific to visualization (i.e., contributing to the \textit{weaponization} stage~\cite{almohannadi2016cyber}).
    
    \item[C3. Long, complex pipeline with interdependent processes.]
    ML-aided visualizations may involve a large number of interdependent processes. 
    This interdependency introduces additional opportunities for adversaries to amplify a cascade of attacks throughout the pipeline.
    For instance, adversaries may be able to create slight influences on Data Processing4VIS so that the subsequent processes will cause critical issues (see \autoref{sec:multiple-process-attack} for a concrete example).
    This cascade of attacks can be even amplified when feedback loops are involved~\cite{sacha2016role}.
    Furthermore, the ML4VIS approach inherits each respective attack surface that exists in ML~\cite{papernot2018sok} and visualization~\cite{mcnutt2020surfacing} pipelines, resulting in an even larger attack surface. 
    This attack surface indicates that ML-aided visualizations might be exposed to more potential threats and could be more complex to defend than cases that only use ML or visualization.
    
    \item[C4. Active involvement of users in the ML processes.]
    Users are actively involved with ML-aided visualizations. They interpret (i.e., VIS Reading) and interactively analyze (i.e., VIS Interaction and User Profiling) the visual content.
    This user involvement provides adversaries opportunities to manipulate users to project attacks or target users as their attack objectives.
    For example, by exploiting the information memorized in NNs, adversaries might reveal users' private information, such as their cognitive ability~\cite{steichen2013useradaptive}.

    \item[C5. Threats on human judgment.]
    Human intervention is vital to avoid critical harm from actions instigated by ML outputs~\cite{heer2019agency,hamon2020robustness,shneiderman2020human}. 
    For example, even if an autonomous car misclassifies a stop sign as a yield sign from an attack, a human driver could still hit the brakes~\cite{trimble2014human}.
    However, with ML4VIS, adversaries may intentionally attack visualizations that require human judgment, making them deceptive and thereby creating situations where human intervention is not as readily feasible (\autoref{sec:substitute-model-attack} demonstrates such situations).
    Although closely related to C4, we separately list this attack surface characteristic. Effective human intervention would be the final defense to protect users from potential harm.
\end{description}

\begin{table*}[t]
  \caption{The summary of concrete adversarial attack examples in \autoref{sec:attack-examples}.}
  \label{table:examples}
  \centering
  \scriptsize
    \begin{tabular}{>{\raggedright}p{1.8in}>{\raggedright}>{\raggedright}p{1.4in}p{2.65in}}
    \toprule
     & ML-aided processes & Capabilities of adversaries\\
    \toprule
    
    One-attribute attack (\autoref{sec:one-attribute-attack})
    & Data Processing4VIS \newline (Parametric DR)
    & Observation of inputs to the ML model and outputs after the visual mapping
    \\
    \midrule
    Attack using a substitute model (\autoref{sec:substitute-model-attack})
    & Data Processing4VIS \newline (Parametric DR)
    & Observation of training data or many newly generated inputs and the corresponding outputs
    \\
    \midrule
    Knowledge-based attack (\autoref{sec:knowledge-based-attack})
    & Data-VIS Mapping \newline (Chart recommendation)
    & Part of the specifications of the ML model
    \\
    \midrule
    Attack referring to gradients (\autoref{sec:gradient-based-attack})
    & Data-VIS Mapping \newline (Chart recommendation)
    & Information on the ML model and gradients
    \\
    \midrule
    Attack propagating across multiple processes (\autoref{sec:multiple-process-attack})
    & Data Processing4VIS \& Data-VIS Mapping 
    & Both the capabilities related to \autoref{sec:substitute-model-attack} and \autoref{sec:gradient-based-attack}
    \\
    \bottomrule
    \end{tabular}
\end{table*}

\section{Concrete Adversarial Attack Examples}
\label{sec:attack-examples}

To showcase possible threats while highlighting the uniqueness of the attack surface, we designed attacks on state-of-the-art ML-aided visualizations~\cite{sainburg2021parametric, wu2022multivision}.
A summary is presented in \autoref{table:examples}. 
Across the attacks, we make different assumptions that cover different levels of the adversaries' capabilities (e.g., black-box attack vs. white-box attack).
The source code used for our attacks is available online~\cite{supp}.

Our attacks are on two representative methods that focus on different parts of the ML4VIS pipeline: (1) parametric UMAP (PUMAP)~\cite{sainburg2021parametric} for Data Processing4VIS and (2) MultiVision~\cite{wu2022multivision} for Data-VIS Mapping.
These methods are selected for three reasons.
First, we aim to cover multiple different processes in the ML4VIS pipeline so we can highlight all of the identified attack surface characteristics (\textbf{C1}--\textbf{5} in \autoref{sec:attack-surface}).
Second, we consider the research influence of these two processes.
Data Processing4VIS is the root process of the ML4VIS pipeline.
As a result, attacks on this process can trickle a cascade of attacks on the subsequent processes.
On the other hand, Data-VIS Mapping is one of the most frequently studied ML4VIS processes~\cite{wang2022ml4vis}.
Lastly, after our extensive search, we found only a limited portion of the published works provide publicly available, executable source codes and datasets~\cite{supp}.
These source codes and datasets are necessary to precisely replicate their ML models.
Consequently, we cover only a few processes in the ML4VIS pipeline---designing attacks on the other processes remains as future work.

\subsection{Attack Target 1: Parametric DR for Data Processing4VIS}

We first provide the background of the attack target.
Nonlinear DR methods, such as t-SNE and UMAP, are commonly used for visual analysis of high-dimensional data~\cite{espadoto2021toward}.
Using NNs, parametric extensions of nonlinear DR methods have been developed (e.g., parametric t-SNE~\cite{lai2022parametric,vandermaaten2009learning} and PUMAP~\cite{sainburg2021parametric}).
Unlike its non-parametric counterpart, the parametric method produces a parametric mapping that projects data instances onto a low-dimensional space. 

\begin{figure}[t]
    \centering
    \includegraphics[width=\linewidth]{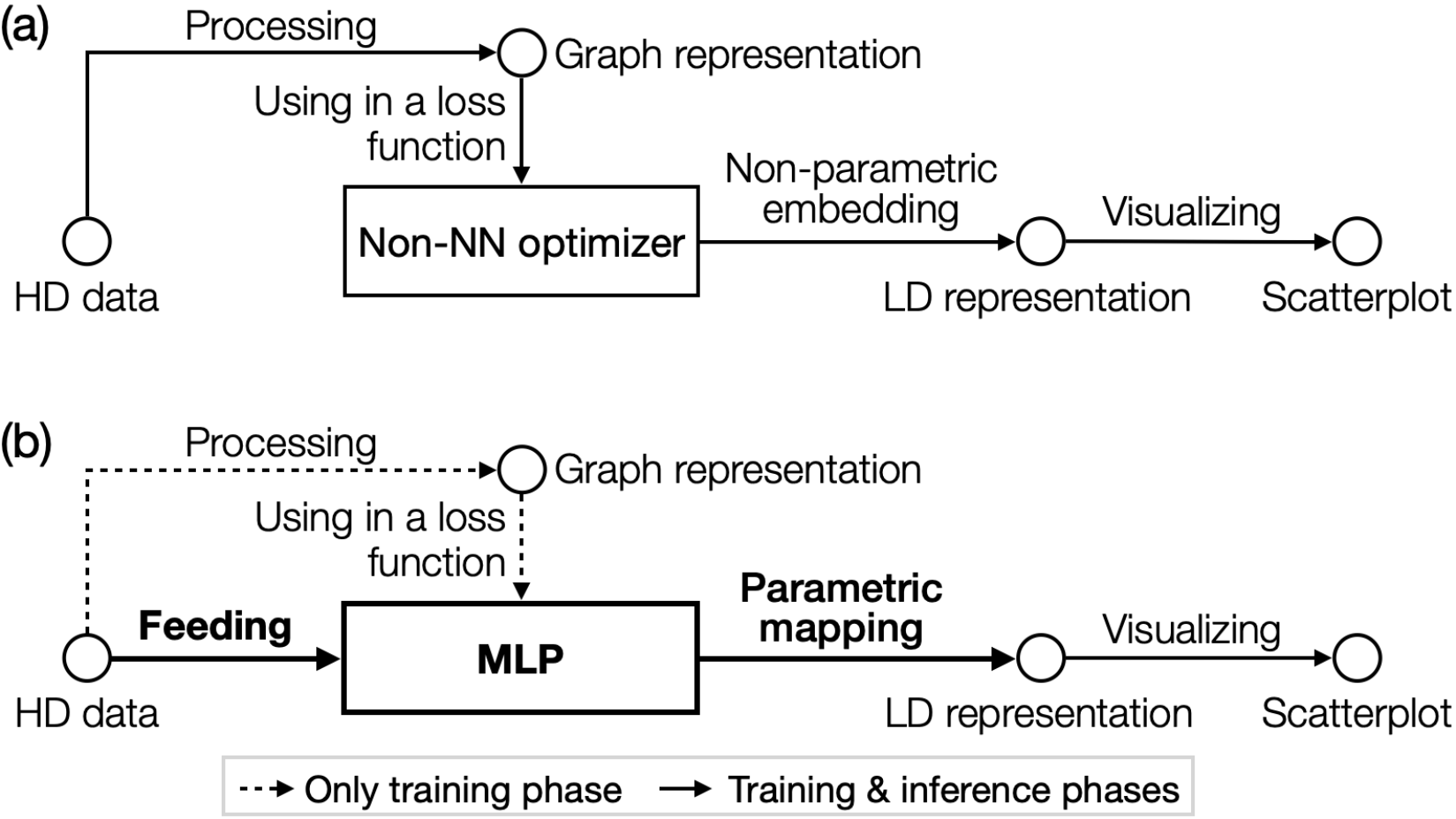}
    \caption{Pipelines for (a) conventional UMAP and (b) parametric UMAP (PUMAP). Unlike UMAP, PUMAP uses a multilayer perceptron (MLP) to learn a parametric mapping from high-dimensional (HD) data to its low-dimensional (LD) representation.
    PUMAP requires the graph representation only for the training phase, as indicated by the dotted lines in (b)}.
    \label{fig:pumap-architecture}
\end{figure}

\autoref{fig:pumap-architecture} compares pipelines employed by UMAP and PUMAP.
Conventional UMAP first constructs a graph representation of the input high-dimensional data.
The subsequent step is an iterative optimization that does not involve NNs. 
Using this optimization, UMAP layouts the instances onto a low-dimensional space (often in 2D) so that instances similar in the graph representation are spatially proximate (refer to \cite{mcinnes2018umap} for details).
There is no direct, numerical connection between high-dimensional data and its low-dimensional representation. 
Thus, UMAP does not provide parametric mapping.
In contrast, PUMAP feeds high-dimensional data to the MLP's input layer, learns the hidden layers' neuron weights for parametric mapping, and produces low-dimensional coordinates from the output layer.
PUMAP only uses the graph representation for a loss function during the training phase.
At the inference phase, PUMAP can directly project new inputs onto the low-dimensional space by using the trained neuron weights. 

This parametric mapping ability is useful for visual analysis, such as when analyzing streaming data.
Since the projection can be performed without rerunning DR methods, parametric DR is computationally efficient and provides visually stable results (e.g., avoiding the arbitrary rotation caused for each update~\cite{fujiwara2020incremental,rauber2016visualizing}).
Parametric DR is actively studied due to this ability and its convenience of NN-based optimizations~\cite{sainburg2021parametric,vandermaaten2009learning,lai2022parametric,hinterreiter2022paradime,zang2024dmtev}.
Also, similar to conventional DR methods, parametric DR can be widely used for visualizations.
Therefore, investigating the vulnerabilities of parametric DR methods is critical to ensure security for ML-aided visualizations.

\subsubsection{Attack Manipulating One Attribute Value}
\label{sec:one-attribute-attack}

\textbf{Attack design.}\quad
For our first attack, using PUMAP as an example, we demonstrate a rudimentary attack that can be performed even with limited adversarial capability. 
This basic example provides evidence of how NN-based parametric DRs are susceptible to maliciously crafted inputs.
We assume that PUMAP is trained with the default setting as introduced by Sainburg et al.~\cite{sainburg2021parametric} (e.g., using an MLP of three 100-neuron hidden layers with rectified linear activation functions, or ReLUs) and that PUMAP outputs are visualized as scatterplots.
In addition, we assume the following: adversaries only have the capability to observe their inputs into the trained PUMAP and the corresponding scatterplots; their goal is to produce misleading scatterplots.

From the assumptions above, we designed a \textit{one-attribute attack}: a black-box attack that manipulates one attribute value of an input instance.
The one-attribute attack is designed based on our observations that PUMAP tends to have highly influential attributes on a learned parametric mapping.
The one-attribute attack consists of two steps.
First, for any given input instance, we observe the change in its low-dimensional coordinate 
before and after adding a small perturbation for each attribute.
We add a small perturbation to hide suspicious behaviors before performing the main attacks.
Although this step is an exhaustive search, we only need to feed $(d{+}1)$ instances at minimum, where $d$ is the number of attributes.
For the second step, we create an adversarial input by adding a relatively large perturbation to the most influential attribute. 
By doing so, we can locate the instance in an arbitrary coordinate in the low-dimensional space.

\begin{figure}[t]
    \centering
    \includegraphics[width=\linewidth]{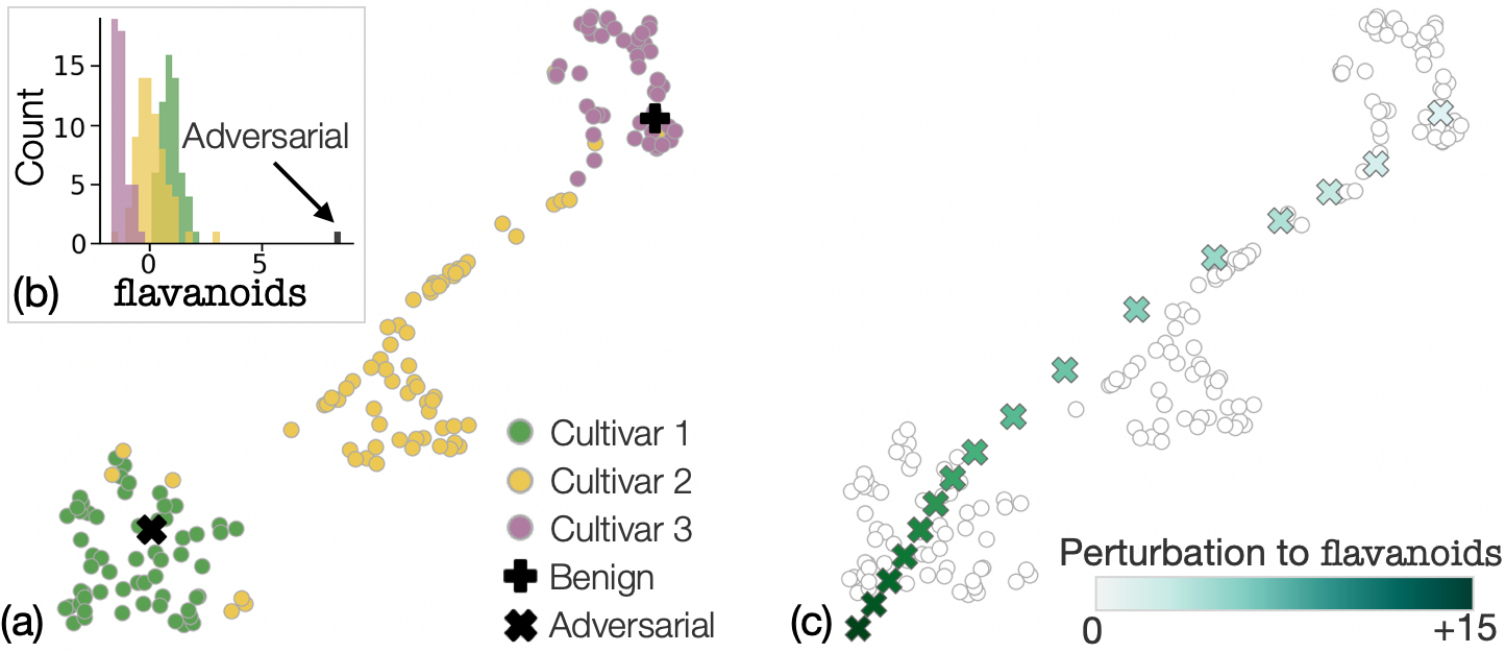}
    \caption{
    Investigation of a one-attribute attack on the visualizations using PUMAP: (a) a scatterplot obtained by applying the default PUMAP to the Wine dataset~\cite{uci_mlr} and the adversarial input; (b) the value distribution of \datatext{flavanoids} for each cultivar and the adversarial input; and (c) the input coordinate migration in response to the perturbations to \datatext{flavanoids}.
    }
    \label{fig:one-attribute-attack}
\end{figure}

\textbf{Attack results and analyses.}\quad
We provide a demonstrative example of the one-attribute attack, using the Wine dataset~\cite{uci_mlr}, which consists of 178 instances, 13 attributes, and 3 cultivar labels.
Since PUMAP is an unsupervised learning method, we use data labels only for the visual encodings to better convey our analysis results.
We first normalize the dataset and then use it to train PUMAP.
\autoref{fig:one-attribute-attack}-a shows the resulting 2D scatterplot, where circles correspond to the 178 instances projected by the trained PUMAP.

Based on the one-attribute attack, \datatext{flavanoids} is identified as the most influential attribute. 
For presentation purposes, we select one \textit{benign} input from \datatext{Cultivar\,3} and then craft an adversarial input by adding a value of 10 into the benign input's \datatext{flavanoids}.
We then project both the benign and adversarial inputs onto \autoref{fig:one-attribute-attack}-a.
The adversarial input is positioned near \datatext{Cultivar\,1} instead of \datatext{Cultivar\,3}. 

\autoref{fig:one-attribute-attack}-b shows a histogram of \datatext{flavanoids} for each cultivar, where \mbox{\datatext{Cultivars\,1}--\datatext{3}} are approximately placed from right to left in order. 
From the histograms, we expect that \datatext{flavanoids} is useful for PUMAP to decide the low-dimensional coordinate of each instance (e.g., the higher value of \datatext{flavanoids}, the more characteristics of \datatext{Cultivar\,1}).
However, as seen in \autoref{fig:one-attribute-attack}-b, although the adversarial input has an extremely high value of \datatext{flavanoids}, the trained PUMAP does not show it as an outlier.
In \autoref{fig:one-attribute-attack}-c, we further investigate the influence of the perturbation strength by incrementing \datatext{flavanoids} from 0 to 15 in step 1. 
As the perturbation strength increases, the adversarial input moves from the top right and passes by all three cultivars. 
By simply changing one attribute, this result highlights how adversaries can manipulate a user's visual perception of which cluster the input instance belongs (e.g., \datatext{Cultivar\,2} when adding 5) while not exposing any extreme or outlying characteristics of the adversarial input.
One can consider that the one-attribute attack is analogous to the one-pixel attack designed for image datasets~\cite{su2019one}.
Although we performed the one-attribute attack on the dataset with 13 attributes, this example is mainly intended to provide a concise demonstration. In reality, the attack may occur for datasets with numerous attributes (e.g., 1,000 attributes).
In such situations, similar to finding the one abnormal pixel in an image, the identification of the one-attribute attack can be non-trivial when humans solely rely on manual or visual inspection of each attribute's data distribution.

We suspect that the observed phenomenon above---the projected coordinate can be controlled even by changing only one attribute value---occurs due to the ``linearity'' of modern NNs. 
Modern NNs utilize close-to-linear activation functions (e.g., ReLU~\cite{dubey2022activation}) and linear multiplications of neuron inputs and weights.
Goodfellow et al.~\cite{goodfellow2015explaining} also hypothesized this linearity characteristic can explain the success of adversarial examples on CNNs. 
As a result, for NN-based parametric DR, some of the attributes could almost be linearly mapped onto one direction of low-dimensional coordinates, as shown in \mbox{\autoref{fig:one-attribute-attack}-c}.
The supplementary material~\cite{supp} exhibits close-to-linear mappings even for different datasets and activation functions.
This inherent issue in NN-based parametric DR should be addressed for more secure use.

\subsubsection{Attack Using a Substitute Model}
\label{sec:substitute-model-attack}

\textbf{Attack design.}\quad
When adversaries can observe various combinations of inputs and the corresponding outputs for parametric DR, they can construct a substitute model that can craft adversarial inputs flexibly and effectively.
Interactive visualization systems showing DR results have a high likelihood of this situation occurring. 
These systems often provide training data information on demand to allow users to examine the DR results~\cite{sacha2016visual}.
Therefore, we assume that it is possible and reasonable that an adversary's goal is to construct a substitute model to the attack target model and generate deceptive visualizations.

\begin{figure}[t]
    \centering
    \includegraphics[width=\linewidth]{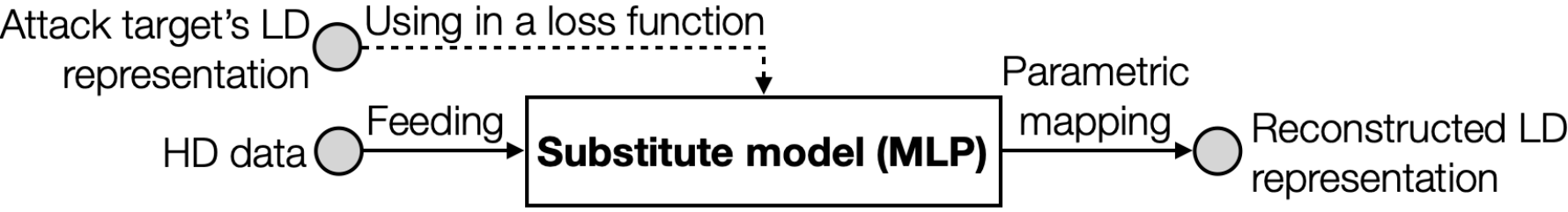}
    \caption{The architecture of the substitute model. By referring to the attack target's low-dimensional (LD) representation, the substitute model learns a parametric mapping that can reconstruct a similar LD representation from the input high-dimensional (HD) data.
    }
    \label{fig:substitute-model}
\end{figure}

\autoref{fig:substitute-model} shows our architecture for a substitute model attack.
The substitute model learns the parameters that produce a nearly identical low-dimensional representation to the one produced by the attack target model. 
This learning is achieved by setting a loss function that minimizes the positional differences (e.g., the sum of pairwise Euclidean distances) between the instances in the low-dimensional representations that are produced by the attack target and substitute model.
To construct this substitute model, adversaries do not need to know how the attack target model's parametric mapping is learned (e.g., the use of PUMAP).
In addition, the substitute model's NN architecture and implementation do not have to be the same as the attack target model's. 
Adversaries need to only have sufficient neurons/parameters to mimic the attack target's parametric mapping. 
Since adversaries can access all information of the substitute model (e.g., gradients), they can efficiently craft adversarial inputs so that the inputs are projected at specific positions in the substitute model's low-dimensional representation. 
Then, adversaries can feed the crafted adversarial inputs to the attack target model, where the inputs would be projected closely to the aimed positions.

\textbf{Attack results and analyses.}\quad 
We demonstrate attacks on the same ML model as \autoref{sec:one-attribute-attack} (i.e., PUMAP trained with the Wine dataset). 
We construct our substitute model with an MLP of three 50-neuron hidden layers, using PyTorch. 
Note: PUMAP employs TensorFlow and an MLP of three 100-neuron hidden layers.

\begin{figure}[t]
    \centering
    \includegraphics[width=\linewidth]{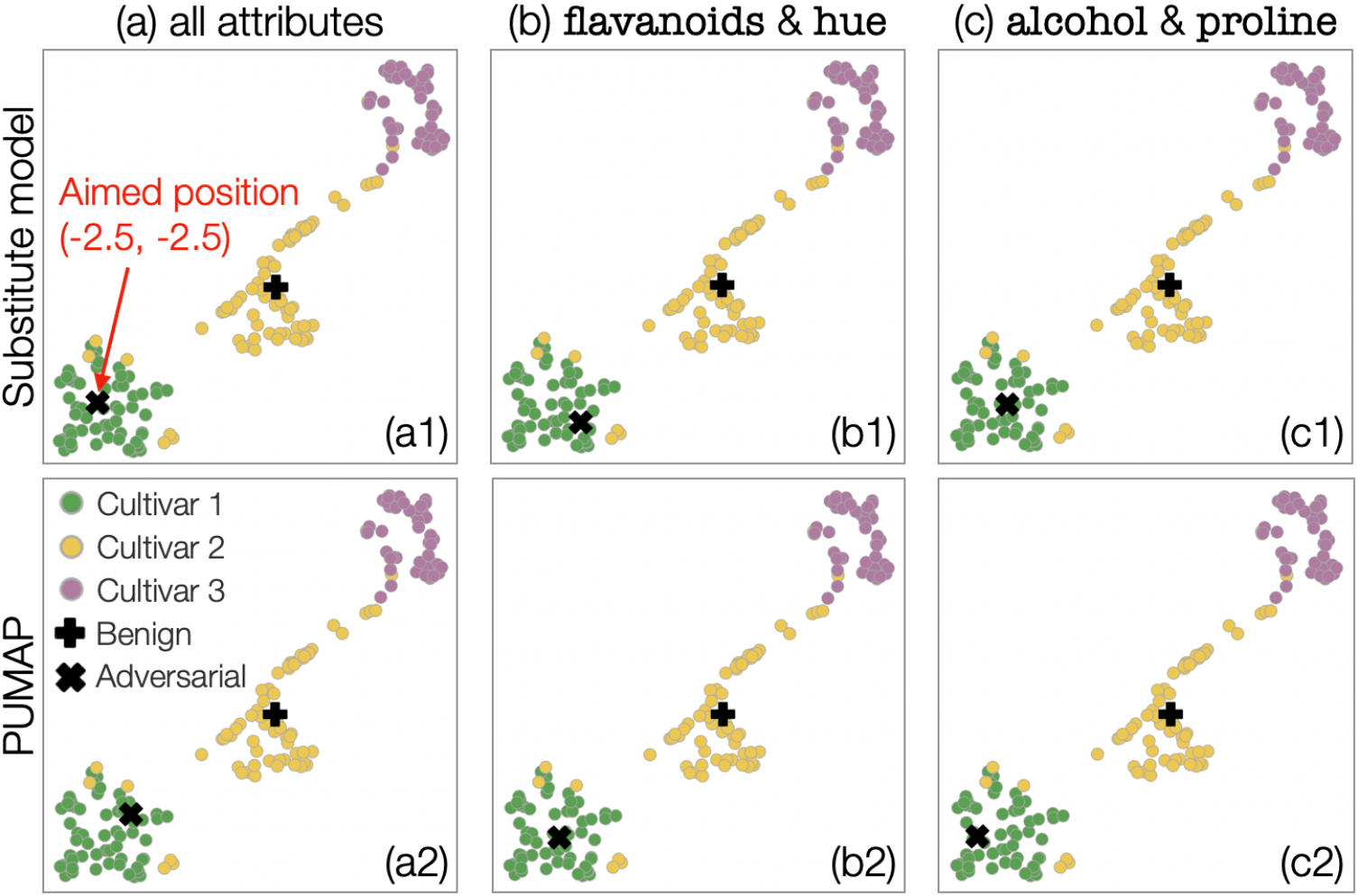}
    \caption{Comparison of adversarial inputs crafted with the substitute model.
    In (a1--c1), all attributes of the benign input, \texttt{flavanoids} \& \texttt{hue}, and \texttt{alcohol} \& \texttt{proline} are respectively manipulated to place the adversarial inputs near the coordinate (-2.5, -2.5). (a2--c2) show the PUMAP results corresponding to (a1--c1).
    }
    \label{fig:substitute-model-adv-inputs}
\end{figure}

We first showcase how to create adversarial inputs.
As shown in \autoref{fig:substitute-model-adv-inputs}-a1, we can create an adversarial input that is projected onto a specific position (e.g., (-2.5, -2.5)) with the substitute model.
The adversarial inputs can be optimized by adjusting the input's attribute values based on their gradients. 
The gradients are related to the error between the aimed and projected positions.
When we feed the crafted adversarial input to PUMAP, as shown in \autoref{fig:substitute-model-adv-inputs}-a2, the adversarial input is placed close to the aimed position.
Due to the linearity of NNs, we expect that similar placement could be achieved without manipulating many attributes. 
For example, two attributes would be enough given how this example's low-dimensional space is in 2D.
To validate this hypothesis, using one arbitrarily selected benign input (in \autoref{fig:substitute-model-adv-inputs}, one instance from \texttt{Cultivar\,2}), we perform this same creation procedure while only manipulating two attribute values of the benign input.
Examples of the produced adversarial inputs are shown in \autoref{fig:substitute-model-adv-inputs}-b and c.
We can see that the adversarial inputs are placed closely to the aimed position in both the substitute model and PUMAP results.

This analysis exposes two critical vulnerabilities for PUMAP as well as for other NN-based parametric DR methods.
First, it can be easy for adversaries to construct a substitute model that has an almost identical parametric mapping to the attack target model's. 
As such, a substitute model can effectively create adversarial inputs without necessarily involving the attack target model.
Second, as discussed in \autoref{sec:one-attribute-attack}, NN-based parametric DR methods are likely to have strong linearity.
By exploiting this linearity characteristic, adversaries can aim to place malicious inputs to desired low-dimensional coordinates by manipulating only a few attributes. 
In the supplementary material~\cite{supp}, we exhibit that this same issue occurs even for three other cases: a PUMAP with a smaller NN, a parametric t-SNE, and different datasets (including a dataset for cancer detection as a life-critical example). 

\begin{figure}[t]
    \centering
    \includegraphics[width=\linewidth]{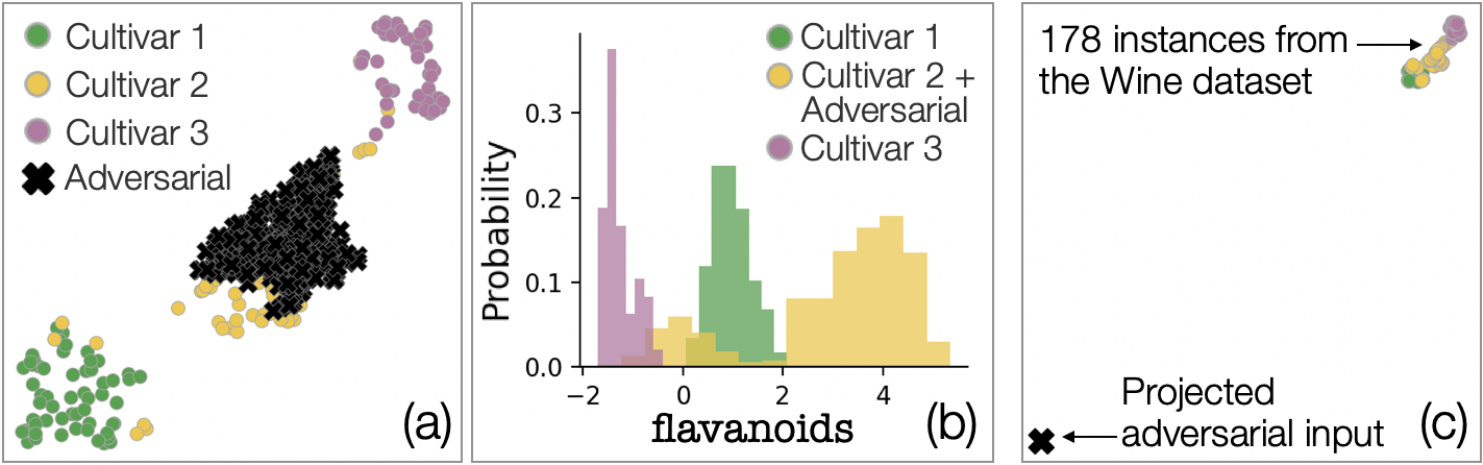}
    \caption{Two attack examples considering the VIS Interaction and VIS Reading processes. In (a), adversaries overwrite the existing visual cluster by manipulating values of \datatext{flavanoids}, \datatext{hue}, and \datatext{ash}.
    If the user interactively constructs three clusters based on (a), the corresponding histogram for each cluster would be similar to (b). In (c), an adversarial input causes extreme scaling by crafting an outlier to conceal the changes around the clusters. 
    }
    \label{fig:pumap-wine-variety}
\end{figure}

From these analysis insights, we can induce that visualizations can be easily obscured, misleading, or deceptive from crafted adversarial inputs.
For example, as shown in \autoref{fig:pumap-wine-variety}-a, adversaries can overwrite the existing visual cluster corresponding to \datatext{Cultivar\,2}.
At the VIS Interaction step, if users aim to identify and understand the clusters from the tainted scatterplot and histograms~\cite{fujiwara2020supporting}, they would fundamentally misunderstand the characteristics of the clusters. 
In \autoref{fig:pumap-wine-variety}-b, a majority of the manipulated cluster (i.e., Cultivar 2) tends to have a higher value of \datatext{flavanoids} than \datatext{Cultivar\,1}. 
In contrast, \datatext{flavanoids} values for the original cluster for \datatext{Cultivar\,2} are mostly in-between \datatext{Cultivar\,1} and \datatext{Cultivar\,3} (see \autoref{fig:one-attribute-attack}-b).
Also, if the Data-VIS Mapping process adjusts the visualization axes based on the value range, adversaries can craft an adversarial input that is projected to be an outlier, as shown in \autoref{fig:pumap-wine-variety}-c.
This radical change of axis scales influences the visual space used for the main region to be much smaller, and thereby affecting the VIS Reading process to disturb observing the main region.
Adversaries may utilize this situation to hide subsequent attacks influencing the main region.

\subsection{Attack Target 2: Chart Recommendation for Data-VIS Mapping}

Chart recommendation is a canonical use of ML for Data-VIS Mapping.
Given input data, chart recommendation systems offer suggestions and generate the appropriate charts.
By using large datasets of collected charts (e.g., \cite{zhou2021table2charts}), researchers applied supervised learning to NNs and developed various chart recommendation systems~\cite{wu2022multivision,zhao2022chartseer}.

\begin{figure}[t]
    \centering
    \includegraphics[width=\linewidth]{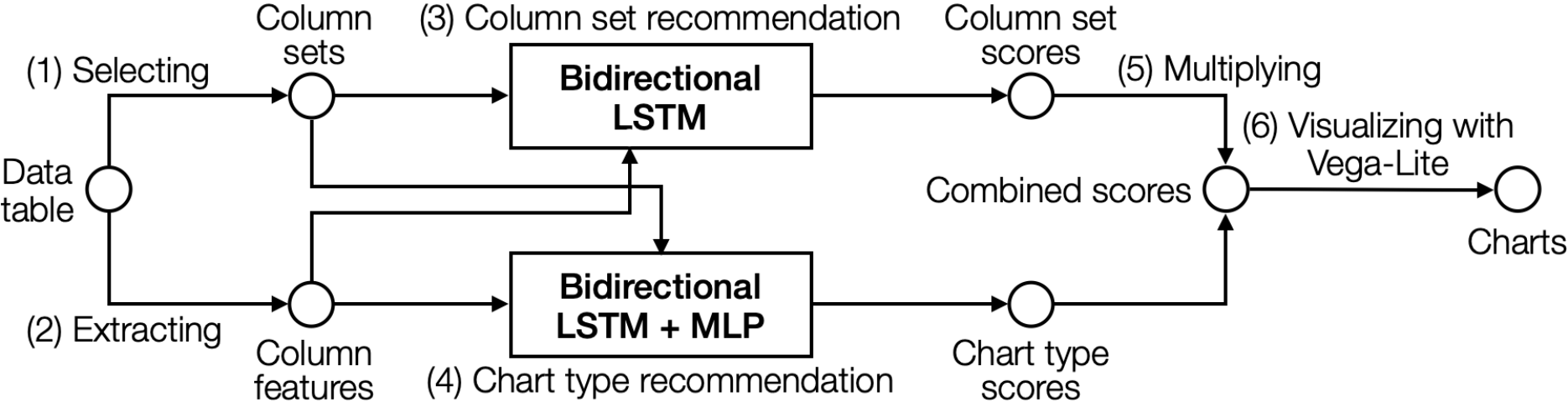}
    \caption{MultiVision pipeline. Two NNs incorporating long short-term memory (LSTM) score the importance of data column sets and the appropriateness of chart types for each of the data column sets. The obtained scores are combined and used to generate recommendations. The recommended charts are then rendered with Vega-Lite~\cite{satyanarayan2017vega}.}
    \label{fig:multivision-architecture}
\end{figure}

We use MultiVision~\cite{wu2022multivision} as a representative example of chart recommendation systems.
Given a data table, MultiVision is designed to rank and recommend multiple charts.
See \autoref{fig:multivision-architecture} for an overview of MultiVision's pipeline for chart recommendation.
The following procedure is applied to both the training and inference phases: 
\begin{enumerate}[topsep=0pt,parsep=0pt,leftmargin=1.5em]
    \item given an input data table, generate sets of data columns for 2D chart generation (e.g., selecting two columns/attributes that would be represented as the scatterplot's $x$- and $y$-axes);
    \item extract various features for each column, such as a word embedding corresponding to the column name, data type (e.g., quantitative, nominal, and temporal), and statistics (e.g., the ratio of negative values, standard deviation, and skewness);
    \item score the sets of columns by feeding the corresponding features to an NN using bidirectional long short-term memory (LSTM); 
    \item select an appropriate chart type (e.g., bar, line, or pie) for each set of columns by using another NN that jointly employs a bidirectional LSTM and an MLP; 
    \item based on the score from the step 3 and the confidence score of the chart type selection from the step 4, rank the recommendation level of each output chart;
    \item visualize the top-recommended charts with Vega-Lite~\cite{satyanarayan2017vega}.
\end{enumerate}

\noindent \\ While chart recommendation systems are useful to help reduce the burden of creating charts for analysts, adversarial attacks can make systems recommend meaningless or deceptive charts and hide important information from the analysts. 
Existing NN-based chart recommendations~\cite{wu2022multivision,zhao2022chartseer,zhou2021table2charts} share similar approaches to MultiVision by extracting data table features as ML inputs and using RNN-related models to interpret the relationships among the set of data columns.

\subsubsection{Knowledge-Based Attack}
\label{sec:knowledge-based-attack}

\textbf{Attack design.}\quad
We perform a simple, but critical attack on Wu et al.'s pre-trained MultiVision~\cite{wu2022multivision}.
Here, we assume that adversaries know or can guess some basic speci\-fications related to the employed ML model.
This assumption fits into cases when adversaries can recognize that the recommendation system uses a similar approach to MultiVision.
Even if adversaries partially know the MultiVision procedure described above, they can strategically craft adversarial inputs. 
For example, to induce misprediction, they can influence data features by manipulating column names or values. 
They can even apply very subtle manipulations to make attacks difficult to notice without a close examination.
To guess the important features for the recommendation, we perform a trial-and-error process to find a subtle change that causes invalid recommendations.

\textbf{Attack results and analyses.}\quad
We show attack results using the Gapminder dataset~\cite{vegadatasets}, which the original authors of MultiVision used for their user study.
We use this dataset to ensure that MultiVision works in the intended manner. 
This dataset consists of 693 instances and 6 attributes with no missing values. 
In the supplementary materials~\cite{supp}, we provide the attack results for another dataset.

\begin{figure}[t]
    \centering
    \includegraphics[width=\linewidth]{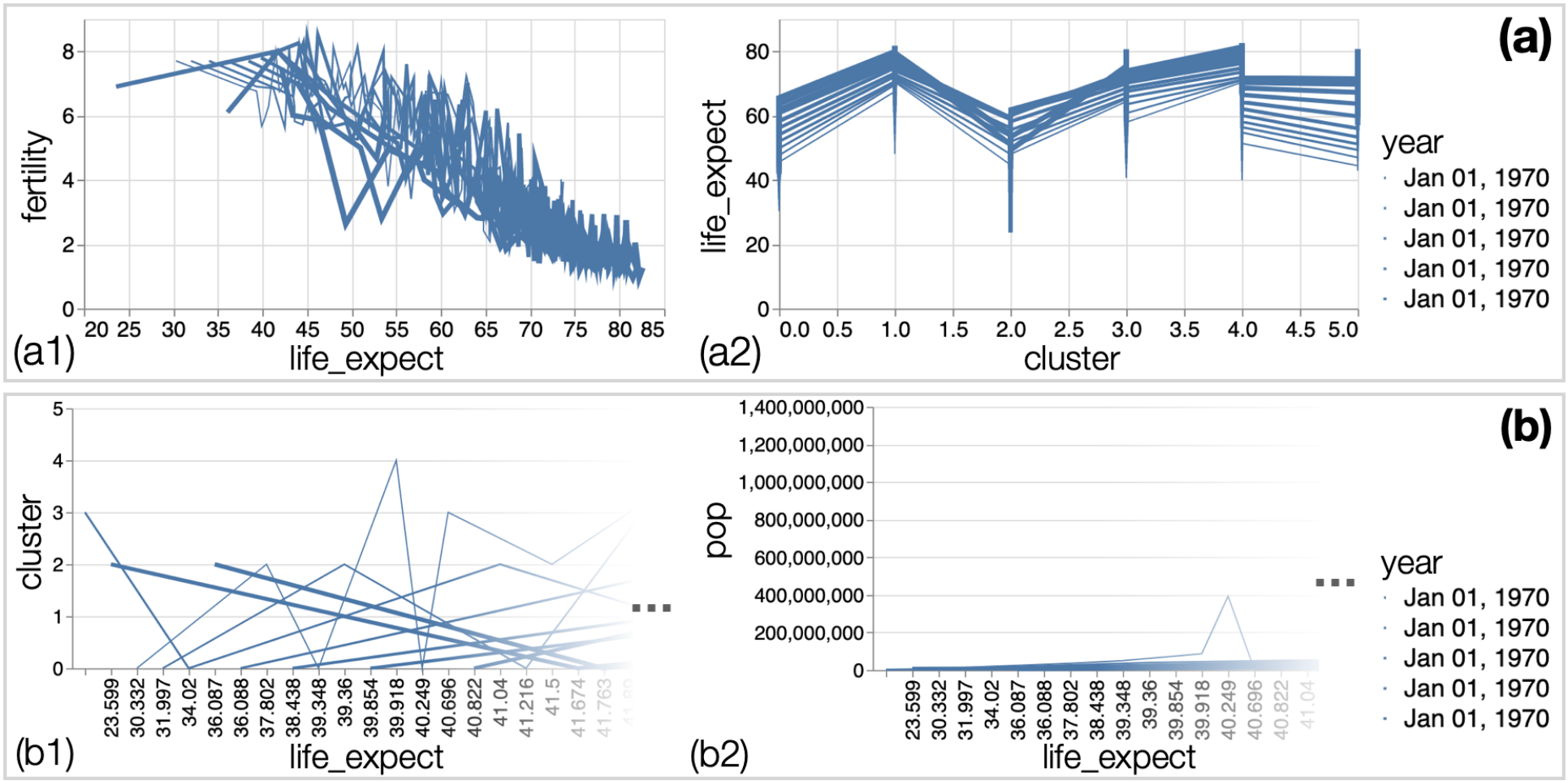}
    \caption{
    The top 2 recommended charts by MultiVision for the Gapminder dataset~\cite{vegadatasets}; (a) before and (b) after the knowledge-based adversarial attack. The charts are positioned in order of the recommendation ranks (i.e., a1: the first, a2: the second).
    In (a), the recommended charts depict useful patterns related to \datatext{life\_expect} (e.g., (a1) shows how \datatext{life\_expect} correlates with \datatext{fertility}).
    In (b), due to the adversarial attack, MultiVision categorizes \datatext{life\_expect} as a nominal attribute, yet still judges \datatext{life\_expect} as an important attribute.
    Consequently, the recommended charts in (b) do not provide meaningful patterns.
    For example, although both (a2) and (b1) use $x$- and $y$-axes to present the relationships between \datatext{life\_expect} and \datatext{cluster}, (b1) does not clearly show the clusters' differences of \datatext{life\_expect} unlike (a2).
    This results in recommending trivial charts. 
    Note that the legend of \datatext{year} is not correct across the charts, while \datatext{year}, \{1955, 1965, ..., 2005\}, is encoded as the line width. This is due to the current implementation of MultiVision (i.e., not due to the attack).
    Moreover, as the charts in b1--b2 are very wide (e.g., 30 times longer than the height), we only show a part of each chart (cf.~\cite{supp} for the full-size charts).
    }
    \label{fig:multivision-gapminder}
\end{figure}

\autoref{fig:multivision-gapminder}-a shows the top 2 recommended charts by MultiVision before our attack. 
MultiVision suggests reviewing the relationships between \datatext{life\_expect} (life expectancy) and other attributes (e.g., \datatext{fertility}) throughout the years. 
From our trials, we decided to generate an adversarial input by replacing one randomly selected value of \datatext{life\_expect} with a blank space (similar to a missing value).
This manipulation leads to the results shown in \autoref{fig:multivision-gapminder}-b.
Though \datatext{life\_expect} is still selected as an important attribute, the resultant charts now do not adequately convey the relationships between \datatext{life\_expect} and other attributes. 
This effect is likely because \datatext{life\_expect} is categorized as a nominal attribute due to the blank space we added. 
Nonetheless, a nominal version of \datatext{life\_expect} is not helpful in understanding the data, and 
MultiVision's susceptibility highlights how it should have selected other attributes for the recommendation charts.

\subsubsection{Attack Referring to Gradients}
\label{sec:gradient-based-attack}

\textbf{Attack design.}\quad
When the chart recommendation system is a white box, adversaries can craft adversarial inputs more efficiently.
MultiVision employs two different NNs to recommend column sets and chart types respectively.
We refer to the gradients for column features toward decreasing the scores corresponding to the highest-ranked chart. 
Then, we manipulate an input data table to change the column features that have large gradient magnitudes.

\begin{figure}[t]
    \centering
    \includegraphics[width=\linewidth]{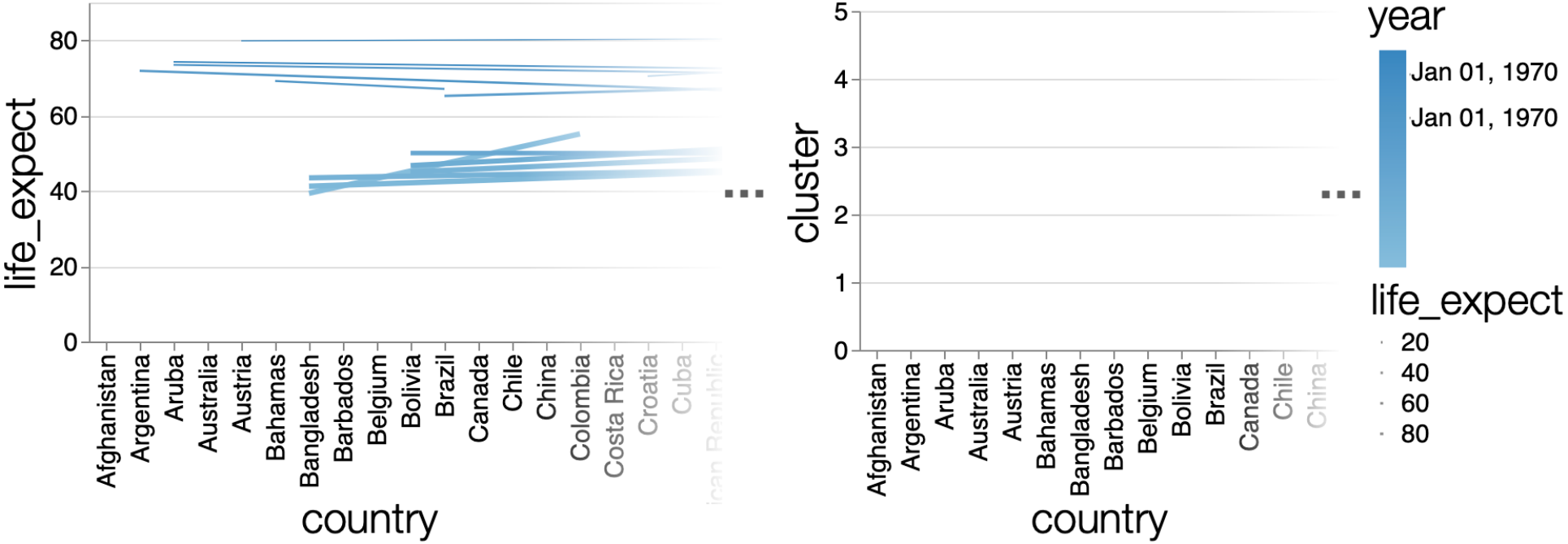}
    \caption{The top 2 charts recommended by MultiVision after shuffling the column order of the Gapminder dataset (cf.~\cite{supp} for the full-size charts).
    }
    \label{fig:multivision-gapminder-whitebox}
\end{figure}

\begin{table}[t]
  \caption{The MultiVision's gradients corresponding to \autoref{fig:multivision-gapminder}-a1 (\texttt{fer.}: \texttt{fertility}; \texttt{LE}: \texttt{life\_expect}). The complete list of features and their meanings is available online~\cite{supp}.}
  \label{table:multivision-gradients}
  \centering
  \scriptsize
  \setlength\tabcolsep{1.8pt}
    \begin{tabular}{r|rrr|rrr}
    \Xhline{2\arrayrulewidth}
    & \multicolumn{3}{c|}{Column set} & \multicolumn{3}{c}{Chart type} \\
    & \multicolumn{1}{c}{\texttt{year}} & \multicolumn{1}{c}{\texttt{fer.}} & \multicolumn{1}{c|}{\texttt{LE}} & \multicolumn{1}{c}{\texttt{year}} & \multicolumn{1}{c}{\texttt{fer.}} & \multicolumn{1}{c}{\texttt{LE}}\\
    \Xhline{2\arrayrulewidth}
    \textbf{\texttt{column\_idx\_normed}} & -0.06 &\textbf{-0.63} & \textbf{-0.48} & -0.04 & \textbf{0.58} & \textbf{-1.00} \\
    \texttt{dataType\_normed} & 0.02 & 0.03 & 0.18 & 0.02 &  0.04 & 0.08\\
    \texttt{aggrPercentFormatted} & -0.05 & -0.10 & -0.22 & 0.02 & -0.11 & -0.14 \\
    ${\cdots}$ & \multicolumn{3}{c|}{$\cdots$} & \multicolumn{3}{c}{$\cdots$} \\
    \Xhline{2\arrayrulewidth}
    \end{tabular}
\end{table}

\textbf{Attack results and analyses.}\quad
We craft an adversarial input for the Gapminder dataset.
As shown in \autoref{fig:multivision-gapminder}-a1, MultiVision originally recommended 
\datatext{life\_expect} ($x$-axis), \datatext{fertility} ($y$-axis), and \datatext{year} (line width) as the column set; line chart as the chart type.
By accessing the NNs used in MultiVision, we extract the aforementioned gradients to disturb this line chart recommendation. 

\autoref{table:multivision-gradients} shows the gradients of the NNs and we list only three out of 96 features used by MultiVision.
We observe that \texttt{column\_idx\_normed} (column index divided by the number of columns) has significantly large gradient magnitudes for both column set and chart type recommendations.
This value indicates that switching the column order introduces a high probability of changing the overall recommendation.
From this observation, we shuffle the order of columns to attack MultiVision, resulting in the recommendation shown in \autoref{fig:multivision-gapminder-whitebox}.
As expected, the recommended charts are radically different from this attack, and the chart no longer seems to be able to support analyzing the dataset.

Bidirectional LSTM considers both forward and backward orders of inputs and theoretically should mitigate the influence of the order of columns. 
However, this attack results clearly highlight that MultiVision is still vulnerable to changes in the column order.
This result draws similar implications to various chart recommendation systems~\cite{wu2022multivision,zhao2022chartseer,zhou2021table2charts} as they also rely on RNN-related models. 
Therefore, the possibility of this vulnerability may be common across these systems.

\subsubsection{Attack Propagating across Multiple Processes}
\label{sec:multiple-process-attack}

\textbf{Attack design.}\quad
This attack focuses on creating adversarial examples for visualizations that utilize NNs for multiple ML4VIS processes.
We consider a system that first performs Data Processing4VIS using PUMAP to derive 2D features from the raw data. 
Then, the subsequent Data-VIS Mapping process employs MultiVision to select the appropriate chart to visualize the 2D features.
We assume that when attacking the target system, adversaries can add new instances to the raw data but cannot directly change a data table input used in MultiVision.
We also assume that adversaries are capable of accessing a pre-trained MultiVision model (e.g., the pre-trained model available from an online repository).
With these two capabilities, adversaries aim to induce misprediction for the chart recommendation.

Similar to \autoref{sec:gradient-based-attack}, we can refer to the gradients of the pre-trained MultiVision models. 
However, when considering the first capability of adversaries, we can only \textit{indirectly} influence attribute values of the input data table. 
For example, we cannot change the column order and column names.
When MultiVision extracts the column features, attribute values (in our case, 2D features from PUMAP) are converted into statistics that are interdependent from each other (e.g., the ratios of negative values and skewness).
Consequently, it is not trivial to associate these statistics' gradients with the data table's attributes. 
Instead, we generate various 2D features with multiple different magnitudes of values and select one that changes the ranks of the chart type or column set scores.
Then, to find an adversarial instance that PUMAP transforms to the selected 2D feature, we build and exploit a substitute model of PUMAP by taking the same approach as \autoref{sec:substitute-model-attack}.

\begin{figure}[tb]
    \centering
    \includegraphics[width=0.95\linewidth]{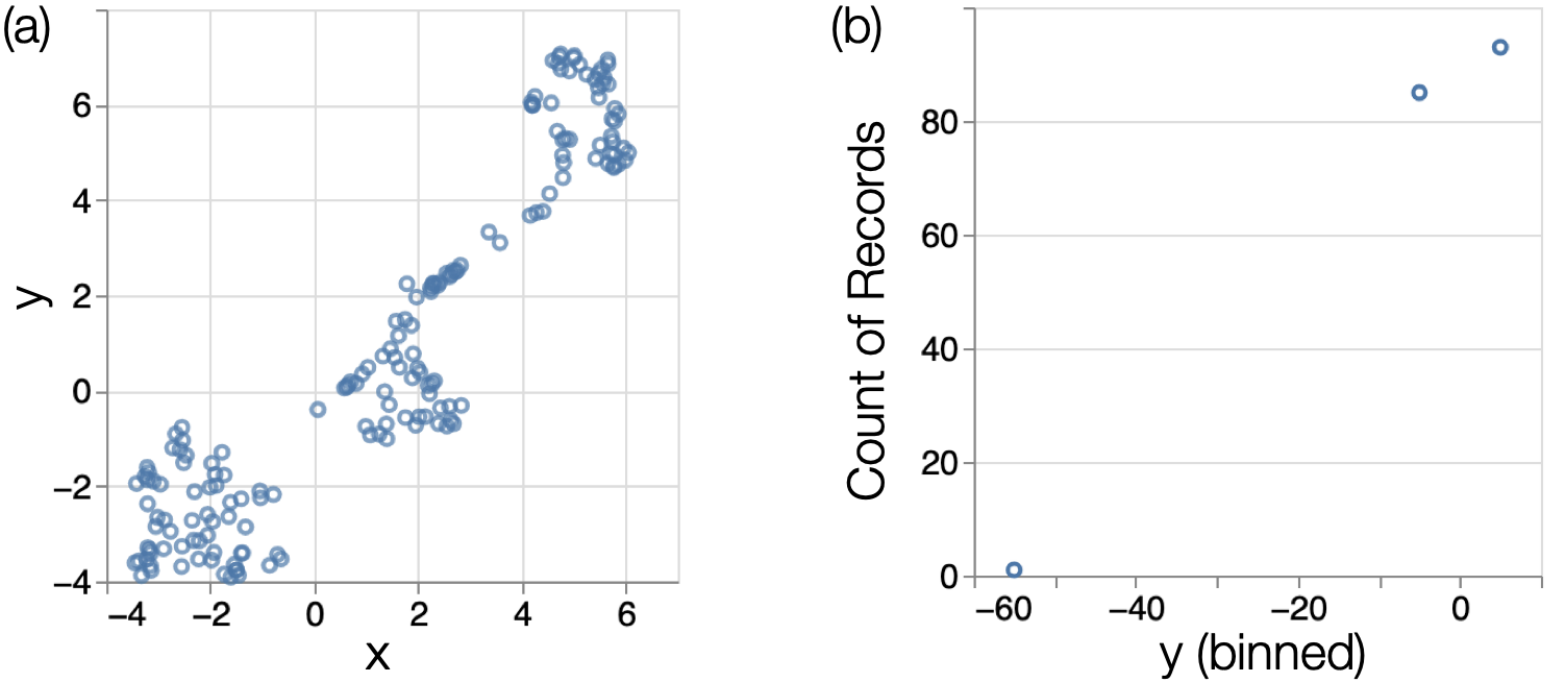}
    \caption{MultiVision's top recommended chart to review the PUMAP result (a) before and (b) after adding an adversarial input.}
    \label{fig:attack-umap-multivision}
\end{figure}

\textbf{Attack results and analyses.}\quad 
We demonstrate an attack that aims to change MultiVision's recommendation by influencing PUMAP's inference.
\autoref{fig:attack-umap-multivision}-a shows MultiVision's top-recommended chart for PUMAP's 2D feature outputs extracted from the Wine dataset.
The recommended chart reasonably visualizes the distribution of features. 
Using the attack design described above, we create and add one adversarial input that is processed as negative 2D feature values with large magnitudes.
Feeding this input to the attack target generates the recommendation shown in \autoref{fig:attack-umap-multivision}-b. 
In this case, the chart is less suitable for visualizing the distribution of 2D features when compared with \autoref{fig:attack-umap-multivision}-a.
This result highlights how MultiVision can drastically change the chart recommendation even with one additional outlier. 

\section{Discussion: Toward Secure Machine Learning-Aided Visualization}
\label{sec:discussion}

We now discuss the implication of the attacks demonstrated in \autoref{sec:attack-examples} as well as highlight potential threats in the real world. 
We then discuss the insights we gained by studying these concrete attacks.
Lastly, we provide a set of suggestions to help further investigate the vulnerabilities of ML-aided visualizations to move toward a more secure use.

\subsection{Discussion on the Performed Attack Examples}

\textbf{Possible critical scenarios in the real world.}\quad
Although we performed the attacks using demonstrative datasets and ML-aided visualizations, these attacks can be easily applied to various real-life scenarios.

Adversaries can directly influence NN models used in various domains.
For example, in the case where NN-based parametric DR is used to monitor a product in real time~\cite{fujiwara2020incremental}, adversaries might notice that the product weight has an influence on the projection. 
This attack would result similarly to the one-attribute attack example.
Then, they might physically control the weight to hide any other abnormal product status.
They may also intentionally cause problematic $x$, $y$-axes scaling in visualizations to conceal any subsequent attacks from analysts.
Another possible scenario is how these adversarial attacks can have political influence. 
For example, parametric DR can be employed to analyze political opinions, such as sentiment analysis in social media posts~\cite{darwish2020unsupervised}. 
Adversaries could find that the projection can be controlled by a post's length and overwrite existing visual clusters by sending many posts of different lengths to hide important opinion differences.

Adversaries can also attack chart recommendations in practical scenarios. 
When a company analyzes its web customer data using chart recommendation services, adversaries (disguising as customers) can make their own data with malicious values to cause problematic recommendations.
As another example, since diagnosing NN models themselves requires analyses from various aspects~\cite{chatzimparmpas2020state}, a company may utilize a chart recommendation service to efficiently detect adversarial attacks on their NNs (i.e., utilizing NN-based chart recommendation to analyze the other NNs).
If a chart recommendation service has vulnerabilities, adversaries can exploit them to hide their attacks on the company's NN model, using a similar approach to the attack propagating across multiple processes we performed.

\textbf{Need of defense strategies for the identified vulnerabilities.}\quad
We identified clear vulnerabilities of PUMAP and MultiVision. 
These vulnerabilities are likely to exist in other NN-based parametric DR methods and chart recommendation systems.
The design and success of the performed attacks are highly related to the unique characteristics of the ML4VIS attack surface, as described in \autoref{sec:attack-surface}.

The attacks on PUMAP utilize ML inputs and outputs exposed by the visualization, exploit the linearity of NNs, and demonstrate cases that can potentially cause user misjudgments.
Developing defense strategies against adversaries requires more research effort. 
For example, concealing the inputs and outputs may reduce the analysis capability or the interpretability of ML-aided visualizations, requiring developers to find a good balance between preserving security and supporting analyses.
The linearity of NNs can be mitigated by employing other activation functions such as the sigmoid. 
However, even these functions have a region that is close to being linear. 
As a result, these functions alone cannot significantly resolve the issues related to the linearity of NNs (see \cite{supp}).  
Through these insights, we may need to design new activation functions to device secure parametric DR methods.

One of the root causes of the success of the attacks on MultiVision could be a mismatch between the employed models (i.e., bidirectional LSTMs) and visualization-specific inputs (i.e., data tables).
Though we want to capture the relationship among table columns, we do not want to place importance on the order of columns when applying ML to data tables.
Although the influence of the column order change can be avoided by shuffling the order during the training phase, a new NN architecture suitable for data tables should essentially be developed.

Also, the attack propagating across multiple ML processes exemplifies a significant security vulnerability for long, complex ML4VIS pipelines.
This attack highlights that not only do we need to investigate each of the ML4VIS processes, but also study how to defend an ML-aided visualization as a whole system.

\subsection{Suggestions for Future Research}
\label{sec:discussion-suggestions}

\textbf{Enhance studies on attacks specific to ML-aided visualizations.}\quad
To develop defense strategies, we first need to analyze possible attacks that exploit the vulnerabilities of ML-aided visualizations.
We expect that various attacks can be specific to ML-aided visualizations based on our observations on their attack surface (\autoref{sec:attack-surface}). 
As a primary step toward a better understanding of ML4VIS' vulnerability, this work demonstrates several critical attacks.
As discussed in \autoref{sec:attack-examples}, only a limited work makes their source code, training and testing datasets, and pre-trained ML models available for the public~\cite{supp}.
In contrast, in the ML field, they are often publicly available given how they are vital to efficiently and accurately study limitations of ML models~\cite{goodfellow2015explaining,su2019one}. 
Thus, we encourage more visualization researchers to make efforts to provide full accounts of data and software publicly, and thereby enable further analyses of ML-aided visualizations.
Also, this immaturity of security studies in ML4VIS indicates our study's limitation. 
The current characterization of the attack surface is based on an abstract-level analysis and may not reflect a variety of possible real attacks (since they are yet unseen).

\textbf{Identify and inform potential vulnerabilities.}\quad
Furthermore, we suggest that researchers routinely identify and publicly inform potential vulnerabilities in their ML-aided visualizations. 
Authors have intimate knowledge of their methods (e.g., designs, algorithms, and datasets).
For example, adding discussions on the vulnerabilities in the respective publications would largely benefit risk assessments.
If we believe the developed visualizations provide significant value (e.g., having a large number of users and deriving highly valuable knowledge)~\cite{vanwijk2006views}, these discussions are crucial in light of potential threats.
Although discussing the vulnerabilities involves a risk of distributing information about attack entries, unsolved critical issues should be reported before the research results are applied to practical applications.
In addition to these individual efforts, there is a growing need to develop methods that systematically evaluate vulnerabilities~\cite{mcnutt2020surfacing}, which would reduce the need for time-consuming manual inspections.

\textbf{Investigate the role of human intervention.}\quad
Lastly, we pose another two open questions: Should we utilize human intervention for detecting adversarial attacks? If so, how?
For pure ML models, when inputs and the corresponding ML predictions have a human-noticeable mismatch (e.g., when a stop sign is recognized as a yield sign), human intervention is useful to avoid harm caused by this mismatch.
However, for ML-aided visualizations, adversaries can aim to create deceptive visualized results to make human intervention unreliable or impossible.
One potential approach to still use human intervention is to provide multiple visualizations that exhibit different responses to the changes in input data. 
For example, for the adversarial inputs crafted in \autoref{sec:substitute-model-attack}, if we constantly show both the scatterplot and histogram (e.g., \autoref{fig:pumap-wine-variety}-a and b), a user would detect the beginning of attacks from the histogram.
Similarly, visual supports designed for the detection of biases in ML model outputs~\cite{borland2021selectionbias} may be useful.
However, approaches that rely on additional visualizations can increase the cognitive load of users, which would be especially problematic when visualizations are used for real-time monitoring purposes.
Further research is required to answer many of these fundamental questions and provide clear and comprehensive guidelines for developers and users.

\section{Conclusion}

We systematically reviewed the vulnerabilities of ML-aided visualizations. 
We described the uniqueness of the attack surface of ML-aided visualizations and demonstrated security risks with five concrete examples of adversarial attacks.
Our results show that more research efforts are needed to address security aspects and advance defense strategies against these adversarial attacks. 

This work also suggests several future research directions, such as investigating diverse adversarial attacks, systematically testing to evaluate the robustness of ML-aided visualizations, and evaluating the human role in defense against adversarial attacks.
In addition to pursuing these directions, we suggest future research to study the interrelationships between security and other closely related topics, such as privacy preservation and trust building~\cite{chatzimparmpas2020state} in ML-aided visualizations.
This work contributes as a stepping stone toward a holistic study for both maximizing benefits and minimizing risks in the use of ML for visualizations.

\backmatter

\bmhead{Supplementary information}

We provide the supplementary materials online~\cite{supp}.
The materials include the source code for the adversarial attack examples in \autoref{sec:attack-examples}; additional experiments related to the attack examples; full-size figures and tables of the attack results; and a list of publications that provide publicly available source code.

\bmhead{Acknowledgments}

The authors wish to thank S. Sandra Bae and Daniel J\"{o}nsson for their assistance in improving the clarity of the paper's content.
This work has been supported in part by the Knut and Alice Wallenberg Foundation through grant KAW 2019.0024 and the ELLIIT environment for strategic research in Sweden.

\bibliography{bibliography}


\end{document}